\documentclass[preprint,prl,superscriptaddress,amsmath,amssymb,aps,floatfix,nofootinbib]{revtex4-2}

\usepackage[pdftex]{graphicx} 
\usepackage{bm} 
\usepackage{color} 
\usepackage{times} 
\usepackage{amsmath} 
\usepackage{amssymb} 
\usepackage{natbib} 
\usepackage{float} 
\usepackage{mathtools}

\begin{document}

\title{Characteristics of fluid-fluid displacement in model mixed-wet porous media: patterns, pressures, and scalings}

% Use letters for affiliations, numbers to show equal authorship (if applicable) and to indicate the corresponding author
\author{Ashkan Irannezhad} \affiliation{Department of Civil Engineering, McMaster University, Hamilton, ON, Canada} 
\author{Bauyrzhan K. Primkulov} \affiliation{Department of Civil and Environmental Engineering, Massachusetts Institute of Technology, Cambridge, MA} 
\author{Ruben Juanes} \affiliation{Department of Civil and Environmental Engineering, Massachusetts Institute of Technology, Cambridge, MA} 
\author{Benzhong Zhao} 
\email{robinzhao@mcmaster.ca} \affiliation{Department of Civil Engineering, McMaster University, Hamilton, ON, Canada}

% \date{\today}
% Single-column figure: width=8.6cm
% Two-column figure: width=17.2cm

\begin{abstract}
	
We study the characteristics of fluid-fluid displacement in simple mixed-wet porous micromodels numerically using a dynamic pore network model. The porous micromodel consists of distinct water-wet and oil-wet regions, whose fractions are systematically varied to yield a variety of displacement patterns over a wide range of capillary numbers. We find that the impact of mixed-wettability is most prominent at low capillary number, and it depends on the complex interplay between wettability fraction and the intrinsic contact angle of the water-wet regions. For example, the fractal dimension of the displacement pattern is a monotonically increasing function of wettability fraction in flow cells with strongly water-wet clusters, but it becomes non-monotonic with respect to wettability fraction in flow cells with weakly water-wet clusters. Additionally, mixed-wettability also manifests itself in the injection-pressure signature, which exhibits fluctuations especially at low wettability fraction. Specifically, preferential filling of water-wet regions leads to reduced effective permeability and higher injection pressure, even at vanishingly small capillary numbers. Finally, we demonstrate that scaling analyses based on a weighted average description of the overall wetting state of the mixed-wet system can effectively capture the variations in observed displacement pattern morphology. 
 
\end{abstract}

\maketitle

\section{Introduction}

Fluid-fluid displacement in porous media is a complex phenomenon with significant practical importance. The behavior of fluid-fluid displacement in porous media impacts various natural and industrial processes such as geological carbon sequestration~\cite{macminn10-jfm1,szulczewski-pnas-2012}, water infiltration into soil~\cite{glass-wrr-1989,cueto-prl-2008}, enhanced oil recovery (EOR)~\cite{orr-science-1984,lake-eor}, and polymer electrolyte membrane (PEM) electrolyzers~\cite{lee-pra-2019,zhao-crps-2021}. Decades of research have illustrated that the displacement process is governed by the competition between capillary and viscous forces as characterized by the capillary number $\text{Ca}$, and the viscosity contrast between the invading and defending fluid as characterized by the viscosity ratio $\mathcal{M}$~\cite{lenormandtouboul88}. In addition, the invading fluid's relative affinity to the solid surface in the presence of the defending fluid (i.e. wettability) as characterized by the contact angle $\theta$ also exerts fundamental control over the flow behavior~\cite{stokes-prl-1986,cieplak-prl-1988,zhao-pnas-2016}. For instance, when the displacing fluid is more wetting to the porous media than the displaced fluid (i.e., imbibition), the fluid-fluid displacement pattern is generally more compact than the opposite case (i.e., drainage)~\cite{holtzman-prl-2015, zhao-pnas-2016,primkulov2018quasistatic, lan2020transitions, yang-labchip-2022}. Today, we have a fairly good understanding of how the interplay between $\text{Ca}$, $\mathcal{M}$, and $\theta$ impact fluid-fluid displacement in porous media with spatially uniform wettability~\cite{primkulov-jfm-2021}. 

In contrast, our knowledge of fluid-fluid displacement in porous media with spatially-heterogeneous wettability (i.e., mixed-wet) is much less complete, despite its common occurrence in natural systems. Mixed-wettability in natural porous systems could originate from the spatial variation in mineral composition and surface roughness. Although reservoirs and aquifers consist primarily of water-wet minerals such as quartz and calcite, they also include other constituent minerals with different wettability states~\cite{ustohal-jch-1998,abdallah2007}. Moreover, water-wet minerals are known to become oil-wet after exposure to polar hydrocarbon components (e.g., asphaltenes) and some micro-organisms~\cite{tweheyo-jpse-1999,ocarroll2005,bultreys2016}. Indeed, high-resolution in-situ contact angle measurements of oil-bearing reservoir rocks have shown mixed-wettability with a wide range of contact angles~\cite{andrew-awr-2014,AlRatrout2018,blunt-jcis-2019}. In the context of geological carbon sequestration, recent experiments have demonstrated that cyclic injection of supercritical $\text{CO}_2$ and brine altered the wettability of a Bentheimer sandstone core, making it less water-wet~\cite{herring-ijggc-2023}. 

The behavior of fluid-fluid displacement in mixed-wet porous media is markedly different from uniform-wet porous media. The significant influence of mixed-wettability on multiphase flow in porous media first emerged in the oil industry, where mixed-wet cores often displayed higher oil displacement efficiency than water-wet cores after waterflooding experiment~\cite{salathiel1973oil,kovscek-aiche-1993}. This behavior is attributed to the existence of interfaces with very low mean curvature and negative Gaussian curvature (i.e., coexisting curvatures with opposite signs in orthogonal directions)~\cite{lin-pre-2019}. Low mean curvature leads to low capillary pressures, while negative Gaussian curvature leads to good connectivity within fluids of the same phase, such that water and oil can flow simultaneously~\cite{scanziani-royal-2020,AlRatrout2018,lin-pre-2019}. In the context of $\text{CO}_2$ storage, it has been observed that $\text{CO}_2$-brine displacement in mixed-wet rocks results in less $\text{CO}_2$ trapping compared to the same process in water-wet rocks~\cite{almenhali-est-2016, almenhali-krevor-est-2016, chang2020impacts}. 

Characterizing the wetting state of porous media with heterogeneous wettability is the first step to understanding fluid-fluid displacement in mixed-wet porous media~\cite{armstrong2021multiscale}. While the wettability of uniform-wet porous media can be described by its contact angle $\theta$, describing the wettability of even the simplest mixed-wet porous media consisting of solids exhibiting just two contact angles requires knowledge of (i) the value of each contact angle, ii) fraction of the porous media covered by solids of each contact angle (i.e., wettability fraction), iii) spatial distribution of the wettability heterogeneity as characterized by the correlation length of clusters with different contact angles as well as their sizes~\cite{guo2020role}.

Due to the various complexities involved in fully characterizing the wetting state of natural mixed-wet porous media, systematic and mechanistic investigation of the impact of mixed-wettability on fluid-fluid displacement in porous media can be greatly aided by the use of simple analog systems with well-controlled geometry and wettability states~\cite{Murison2014, Hiller2019, Geistlinger2021, irannezhad2023fluid}. \citet{Geistlinger2021} performed waterflooding experiments in an air-filled mixed-wet micromodel, which was fabricated by mixing water-wet ($\theta=0^{\circ}$) and oil-wet ($\theta=100^{\circ}$) glass beads in a cylindrical container. They observed a decrease in residual trapping as the fraction of water-wet beads increased from 30\% to 70\%. \citet{Murison2014} conducted waterflooding experiments in oil-filled bead columns where half of the beads' area was water-wet ($\theta=20^{\circ}$) while the other half was oil-wet ($\theta=130^{\circ}$). They observed smoothing of the fluid-fluid displacement front as the correlation length of different wettability clusters decreased. More recently, \citet{irannezhad2023fluid} studied the radial displacement of oil by water in mixed-wet microfluidic flow cells patterned with cylindrical posts. The bulk of the flow cell was oil-wet ($\theta=120^{\circ}$), and mixed-wettability was introduced by placing discrete water-wet clusters that were either weakly water-wet ($\theta=60^{\circ}$) or strongly water-wet ($\theta=30^{\circ}$). The experiments revealed surprising displacement patterns that arise as a result of mixed-wettability~---~the invading water preferentially fills strongly water-wet clusters but encircles weakly water-wet clusters instead. This counter-intuitive finding was attributed to the fluid-fluid interface configuration at mixed-wet pores, which resemble S-shaped saddles with mean curvatures close to zero. 

While well-designed analog experiments have contributed to our mechanistic understanding of fluid-fluid displacement in mixed-wet porous media, they remain prohibitively expensive to sweep a wide range of the relevant parameter spaces. Pore-scale modelling is a useful alternative in this endeavor, since it has become increasingly more predictive in recent years~\cite{zhao-pnas-2016}. \citet{bakhshian-awr-2019} investigated the effect of wettability heterogeneity on the flow of supercritical $\text{CO}_2$ and brine in a $1.2$~mm$^3$ digital rock sample using lattice Boltzmann (LB) simulations. Specifically, they increased the fraction of $\text{CO}_2$-wet portions of the simulation domain from 10\% to 50\% and observed more residual trapping of $\text{CO}_2$. However, LB simulations remain computationally challenging and expensive~\cite{zhao-pnas-2019}. In contrast, pore-network models are attractive due to their intuitive nature and relatively low computational cost~\cite{blunt-cocis-2001}, and they have been applied in a recent study to extend the classic Lenormand's diagram to include the impact of wettability~\cite{primkulov-jfm-2021}. 

Here, we employ a dynamic pore network model to investigate the impact of mixed-wettability on fluid-fluid displacement in simple model porous media consisting of water-wet and oil-wet regions with distinct contact angles. For each contact-angle pair, we systematically vary the wettability fraction and study the displacement pattern over a wide range of $\text{Ca}$. We find that the displacement pattern is controlled by the interplay between $\text{Ca}$, wettability fraction, and intrinsic wettability of the water-wet regions, leading to complex behaviors. Furthermore, mixed-wettability induces fluctuations in the injection pressure, whose magnitude has a large viscous component, even at vanishingly small $\text{Ca}$. Finally, we demonstrate that scaling analyses based on a simple, weighted average description of the overall wettability of the model mixed-wet porous media can effectively predict the finger width of the displacement pattern. 

\section{Method}

We design a microfluidic porous media by placing $\sim16,000$ cylindrical posts on an irregular triangular lattice. The lattice is generated inside a 5-inch diameter circle using the \texttt{pdemesh} tool in MATLAB. We assign the radius of each post to be 47\% of the distance between its center and the nearest neighboring post's center. In our system, the radius of the posts follow a Gaussian-like distribution that ranges from $110~\mu$m to $850~\mu$m with a median value of $340~\mu$m. The pore throat sizes follow a lognormal-like distribution that ranges from $50~\mu$m to $700~\mu$m with a median value of $200~\mu$m. We make the posts' height $h=200~\mu$m to match the median pore throat size.

\begin{figure*}
	[htp] \centering 
	\includegraphics[width=16.8cm]{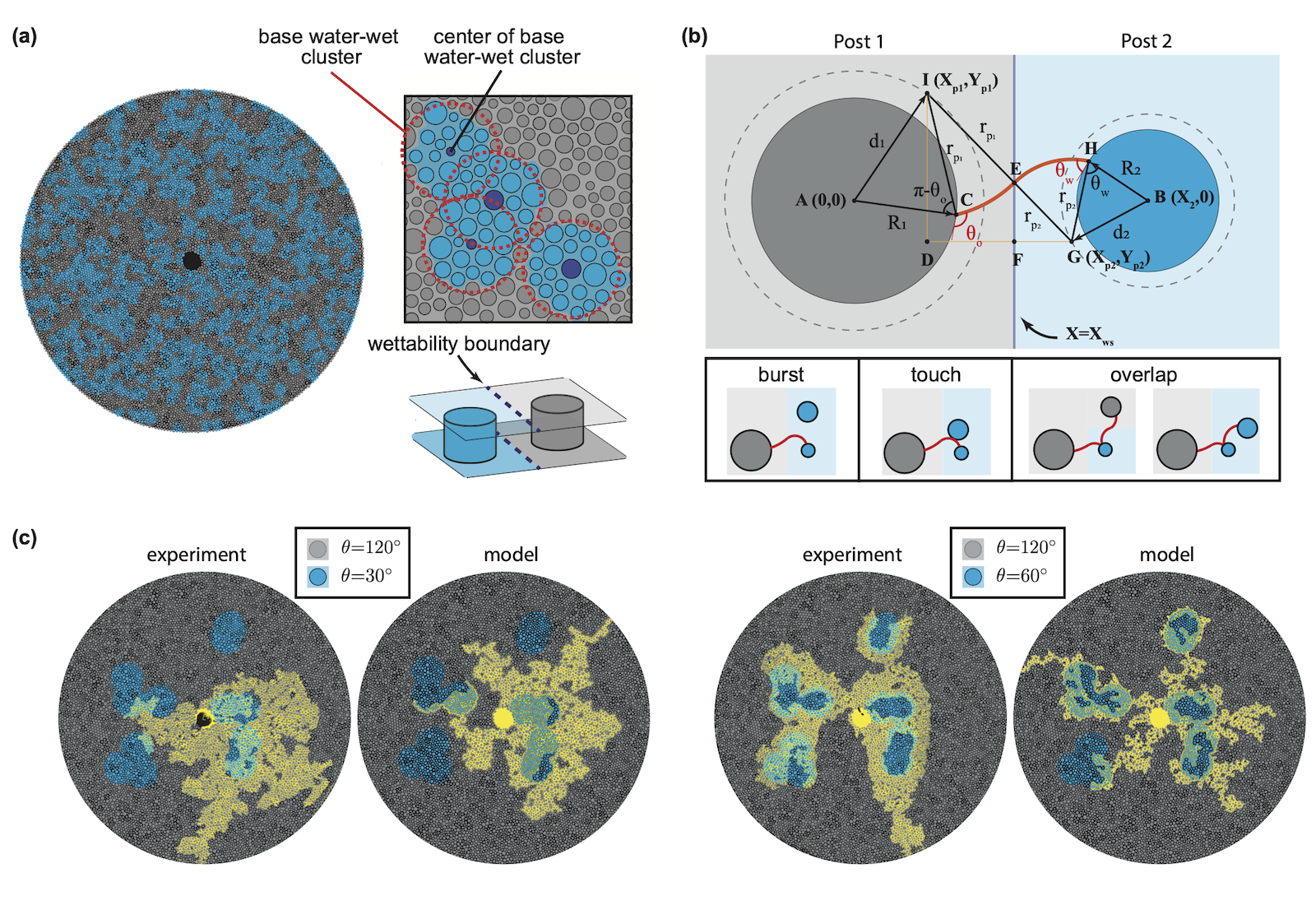} 
	\caption{(a) We develop a dynamic pore network model to simulate fluid-fluid displacement through analog porous media with spatially heterogeneous wettability (i.e., mixed-wet). The flow geometry is radial and quasi-two-dimensional, consisting of $\sim16,000$ cylindrical posts. We achieve mixed-wettability by placing base water-wet clusters (blue regions) that span 5 posts in a domain that is otherwise oil-wet (gray regions). The base water-wet clusters merge and form larger clusters (top inset). The water-wet and oil-wet regions are separated by sharp boundaries (bottom inset). (b) Fluid-fluid interface at a mixed-wet pore throat resembles an S-shaped saddle in three dimensions~\cite{irannezhad2023fluid}. Our model explicitly calculates the critical pressures of pore-scale instabilities and advances the fluid-fluid interface when either a burst, touch, or overlap event occurs. (c) Our pore network model captures the nuanced and complex behavior of multiphase flow in mixed-wet porous media~\cite{irannezhad2023fluid}. The invading water (yellow) preferentially fills the water-wet clusters with contact angle $\theta=30^{\circ}$ (left panel), but encircles the water-wet clusters with $\theta=60^{\circ}$ (right panel). The oil-wet clusters have $\theta=120^{\circ}$ in both cases. Experiments and simulations are conducted at $\text{Ca}=1 \times 10^{-4}$~\cite{irannezhad2023fluid}.} 
\label{fig:mixed-wet_model} 
\end{figure*}

We implement mixed-wettability by placing water-wet clusters in a simulation domain that is otherwise oil-wet. Specifically, we randomly assign $N$ posts as the seeds for generating the water-wet clusters. At each seed post, we draw a circle with a radius of 5 median posts and make all posts and surfaces within the circle water-wet. Overlapping circles merge and form a single cluster~(Fig.~\ref{fig:mixed-wet_model}a). Increasing $N$ increases the wettability fraction $f_\text{w}$, which is defined as the number of water-wet posts to the total number of posts. In this model, oil-wet and water-wet regions are separated by sharp boundaries connecting the adjacent pores centers (Fig.~\ref{fig:mixed-wet_model}a top inset). The top and bottom surfaces of the mixed-wet pores also have spatially heterogeneous wettability (Fig.~\ref{fig:mixed-wet_model}a bottom inset).

We apply a dynamic pore-network model to simulate fluid-fluid displacement in mixed-wet microfluidic cells. The model was first introduced in~\citet{irannezhad2023fluid}, but we describe its details for the first time here. Our mixed-wet model is based on the model of fluid-fluid displacement in uniform-wet porous media developed by~\citet{Primkulov2019}. The model establishes an analogy between multiphase flow in porous media and electrical current in a circuit with fixed resistors and moving capacitors. This analogy enables the model to capture the effects of capillary pressure (at pore throats where fluid-fluid interfaces exist) and viscous pressure drop (at all pore throats) simultaneously throughout the network. Adding wettability effects to the model requires an understanding of invasion mechanisms at the pore-scale, where fluid-fluid interfaces advance as invading fluid pressure increases. This advancement eventually leads to the invasion of the adjacent pore through one of the three possible types of invasion events~\cite{cieplak-prl-1988, cieplakrobbins90}: i) \emph{Burst}, equivalent to Haines jump~\cite{haines-jas-1930}, occurs when there is no stable configuration for a fluid-fluid interface connecting two posts. Thus, the interface suddenly invades the adjacent pore and splits into two interfaces; ii) \emph{Touch} occurs when the fluid-fluid interface connecting two posts touches a third post's body. Similar to the burst event, the fluid-fluid interface splits into two interfaces; and iii) \emph{Overlap} occurs when two neighboring fluid-fluid interfaces meet on a common post. In this case, the two interfaces merge and form a single, wider interface spanning the neighboring posts. The configuration of any given fluid-fluid interface can be described by its in-plane and out-of-plane curvatures. 

In a quasi-2D, uniform-wet microfluidic cell, the interface's out-of-plane radius of curvature $r_\text{out}$ is constant and equal to $h/(2\textrm{cos}\theta)$, while its in-plane radius of curvature $r_\text{in}$ decreases as the interface advances through the pore throat. Therefore, capillary pressure at the interface gradually increases as the interface advances, until the interface undergoes an invasion event (i.e., burst, touch, or overlap). The capillary pressure associated with pore invasion is termed the critical capillary pressure, and it is given by $p_c^\text{crit}={\sigma/r_\text{out}}+\text{min}\left(\sigma/{r_\text{in}^\text{burst}},\sigma/{r_\text{in}^\text{touch}}, \sigma/{r_\text{in}^\text{overlap}}\right)$, where $\sigma$ is the interfacial tension between the two fluids, and ${r_\text{in}^\text{burst}}$, ${r_\text{in}^\text{touch}}$, ${r_\text{in}^\text{overlap}}$ are the in-plane radii of curvatures corresponding to burst, touch and overlap events, respectively. These in-plane radii of curvatures for pore invasion mechanisms in a uniform-wet microfluidic cell have been derived in~\citet{primkulov2018quasistatic}.

The fluid-fluid interface configuration at a mixed-wet pore throat is more complex. In particular, the interface has been observed to be saddle-shaped~\cite{lin-pre-2019}. Additionally, in-situ measurements of fluid-fluid interfaces in carbonate rocks have revealed that their mean curvatures are noticeably lower in mixed-wet rocks compared to similar water-wet rocks~\cite{lin-pre-2019, Armstrong2012}. Recent experiments in mixed-wet microfluidics provided high-resolution visualization of the fluid-fluid interface between two posts of contrasting wettabilities, which resembles an S-shaped saddle in 3D with mean curvatures close to zero~\cite{irannezhad2023fluid}. 

The in-plane view of a typical S-shaped fluid-fluid interface at a mixed-wet pore throat is shown in Fig.~\ref{fig:mixed-wet_model}b. Post 1 is an oil-wet post with contact angle $\theta_\text{o}$, radius $R_1$, and center coordinate A $(0, 0)$, while post 2 is a water-wet post with contact angle $\theta_\text{w}$, radius $R_2$, and center coordinate B $(X_2, 0)$. We approximate the S-shaped interface as two circular arcs ($\overset{\huge\frown}{CE}$ and $\overset{\huge\frown}{EH}$) that smoothly connect at the wettability boundary. The centers of $\overset{\huge\frown}{CE}$ and $\overset{\huge\frown}{EH}$ are located at I ($X_\text{p1}$, $Y_\text{p1}$) and G ($X_\text{p2}$, $Y_\text{p2}$), while their radii are given by $r_\text{p1}$ and $r_\text{p2}$, respectively. We denote the distance between points A and I as $d_1$, which is given by the law of cosines at the triangle $\triangle$AIC as 
\begin{equation}
	d_1 =\sqrt {R_{1}^2+r_\text{p1}^2 - 2R_{1}r_\text{p1} \cos (\pi-\theta_\text{o})}. 
\label{eq:one} \end{equation}
Similarly, the distance $d_2$ between points B and H is given by 
\begin{equation}
	d_2=\sqrt {R_{2}^2+r_\text{p2}^2 - 2R_{2}r_\text{p2} \cos \theta_\text{w}}. 
\label{eq:two} \end{equation}
As the fluid-fluid interface advances through the pore throat, point I and point G each traces a circle described by 
\begin{subequations}
	\begin{align}
		X_\text{p1}^2+Y_\text{p1}^2 = d_1^2, \\
		(X_\text{p2}-X_2)^2+Y_\text{p2}^2 = d_2^2. 
	\end{align}
	\label{eq:three} 
\end{subequations}
Since $\overset{\huge\frown}{CE}$ and $\overset{\huge\frown}{EH}$ smoothly join at point E, we construct the right triangle $\triangle$IGD and find the distance $d_3$ between points I and G 
\begin{equation}
	d_3=r_\text{p1}+r_\text{p2}=\sqrt {(X_\text{p2}-X_\text{p1})^2+(Y_\text{p1}-Y_\text{p2})^2}. 
\label{eq:four} 
\end{equation}
Furthermore, the right triangles $\triangle$IGD and $\triangle$EGF are similar triangles, which leads to 
\begin{equation}
	\frac{r_\text{p2}}{r_\text{p1}+r_\text{p2}} = \frac{X_\text{p2}-X_\text{wb}}{X_\text{p2}-X_\text{p1}}, 
\label{eq:five} \end{equation}
where $X_\text{wb}$ is the x-coordinate of the wettability boundary. 

Finally, we note that the mean curvature of the fluid-fluid interface in the oil-wet region must equal that in the water-wet region at equilibrium, since capillary pressure is constant along the entire interface. Therefore, 
\begin{equation}
	\frac{1}{r_\text{p1}}+\frac{2\cos\theta_\text{o}}{h}=-\frac{1}{r_\text{p2}} + \frac{2 \cos \theta_\text{w}}{h}. 
\label{eq:six} \end{equation}
The shape of the fluid-fluid interface at the mixed-wet pore throat is fully described by $X_\text{p1}$, $Y_\text{p1}$, $X_\text{p2}$, $Y_\text{p2}$, $r_\text{p1}$, and $r_\text{p2}$. For a given $r_\text{p1}$, we first solve for $r_\text{p2}$ (Eq.~\eqref{eq:six}). We then solve for $d_1$ and $d_2$ (Eqs.~\eqref{eq:one}-\eqref{eq:two}), followed by solving for $X_\text{p1}$, $Y_\text{p1}$, $X_\text{p2}$, and $Y_\text{p2}$ (Eqs.~\eqref{eq:three}-\eqref{eq:five}). 

We follow the algorithm below to find the critical capillary pressures associated with burst, touch, and overlap invasion events. Starting with a large $r_\text{p1}$, we solve Eqs. \eqref{eq:one}-\eqref{eq:six}. If no valid solution is found, we decrease $r_\text{p1}$ and repeat the process until a valid fluid-fluid interface is found. We then depict the interface and check if touch or overlap will occur. Touch occurs when the depicted interface intersects a neighboring third post, while overlap occurs when the depicted interface intersects an adjacent interface~\cite{cieplak-prl-1988,cieplakrobbins90}. We successively decrease $r_\text{p1}$ until no valid solution can be found, which corresponds to the point when burst occurs. Similar to the homogeneous-wet case, the critical capillary pressure for pore invasion is given by 
\begin{equation}
	p_c^\text{crit}={\frac{2\sigma\cos\theta_\text{o}}{h}}+\text{min}\left(\sigma/{r_\text{p1}^\text{burst}},\sigma/{r_\text{p1}^\text{touch}}, \sigma/{r_\text{p1}^\text{overlap}}\right). 
\label{eq:seven} 
\end{equation}
Our model captures the effects of viscous pressure drop across the system by solving the equations of Poiseuille's law and mass conservation for the network of pore throats. We impose constant flow boundary condition at the inlet pore throats and constant pressure boundary condition at the outlet pores along the perimeter. We assign the viscosities of the defending fluid and the invading fluid to be $\mu_{\rm{def}}=50$~mPa$\cdot$s and $\mu_{\rm{inv}}=0.99$~mPa$\cdot$s, respectively, and the interfacial tension between the fluids to be $\sigma=13$~mN/m. These fluid-fluid properties correspond to the microfluidic experiments of~\citet{irannezhad2023fluid}, which enables direct comparison between the model predictions and the experiments. Specifically, the model captures the salient behaviors of the mixed-wet microfluidic experiment – the invading fluid preferentially fills the water-wet clusters with contact angle $\theta=30^{\circ}$, but encircles the weakly water-wet clusters with contact angle $\theta=60^{\circ}$ instead (Fig.~\ref{fig:mixed-wet_model}c). The oil-wet regions of the flow cell have a contact angle of $\theta=120^{\circ}$ in both cases. 

\section{Results} 
\begin{figure*}
	[ht] \centering 
	\includegraphics[width=16.8cm]{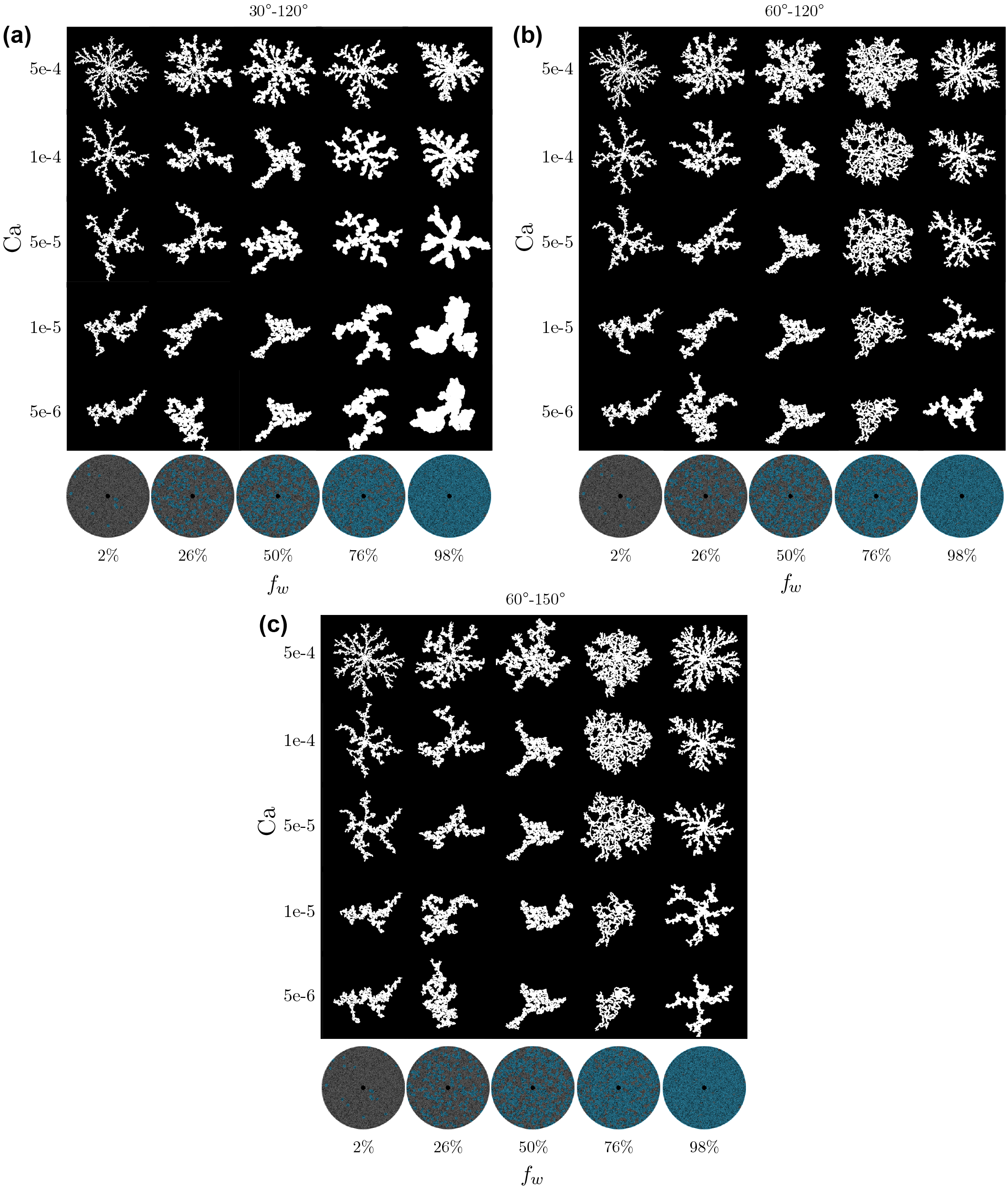} 
	\caption{Phase diagrams of the invading fluid morphology at breakthrough for different wettability fractions (left to right: $2\%$, $26\%$, $50\%$, $76\%$, $98\%$) and capillary numbers (top to bottom: $\text{Ca}=5 \times 10^{-4}$, $1 \times 10^{-4}$, $5 \times 10^{-5}$, $1 \times 10^{-5}$, $5 \times 10^{-6}$). (a) Phase diagram corresponding to mixed-wet porous media with contact angle pairs of $30^{\circ}$-$120^{\circ}$. (b) Phase diagram corresponding to mixed-wet porous media with contact angle pairs of $60^{\circ}$-$120^{\circ}$. (c) Phase diagram corresponding to mixed-wet porous media with contact angle pairs of $60^{\circ}$-$150^{\circ}$.} 
\label{fig:phase_diagram} 
\end{figure*}

\begin{figure*}
	[ht] \centering 
	\includegraphics[width=16.8cm]{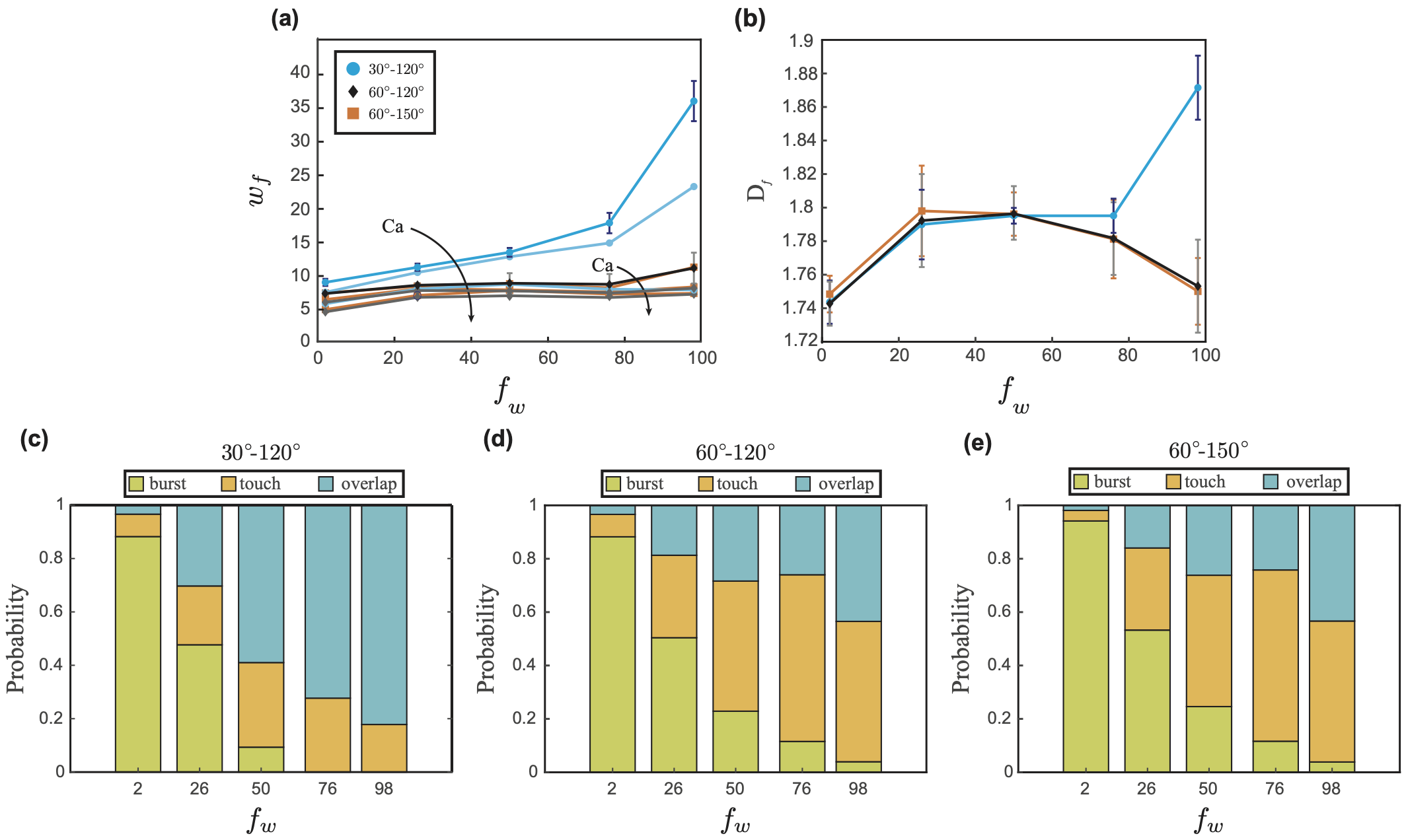} 
	\caption{Quantitative measures of the displacement patterns. (a) Normalized finger width and (b) fractal dimension in mixed-wet domains with contact angle pairs of $30^{\circ}$--$120^{\circ}$ (blue circles), $60^{\circ}$--$120^{\circ}$ (black diamonds) and $60^{\circ}$--$150^{\circ}$ (orange squares). The data at the lowest capillary number ($\text{Ca}=5\times10^{-6}$) is highlighted. Distribution of burst, touch, and overlap events as a function of $f_w$ in mixed-wet domains with contact angle pairs of (c) $30^{\circ}$--$120^{\circ}$, (d) $60^{\circ}$--$120^{\circ}$ and (e) $60^{\circ}$--$150^{\circ}$.} 
\label{fig:plots} 
\end{figure*}

\noindent\textbf{Phase diagram.} We apply the dynamic pore-network model to investigate fluid-fluid displacement in simple mixed-wet microfluidics. Specifically, each simulation domain consists of water-wet regions with contact angle $\theta_\text{w}$ and oil-wet regions with contact angle $\theta_\text{o}$. We consider three different contact angle pairs~---~$\theta_\text{w}$-$\theta_\text{o}$$=30^{\circ}$-$120^{\circ}$, $60^{\circ}$-$120^{\circ}$, $60^{\circ}$-$150^{\circ}$. Our systematic investigation includes varying the wettability fraction ($f_\text{w}=2\%$, $26\%$, $50\%$, $76\%$, $98\%$) and capillary number ($\text{Ca}=5 \times 10^{-4}$, $1 \times 10^{-4}$, $5 \times 10^{-5}$, $1 \times 10^{-5}$, $5 \times 10^{-6}$) over a wide range of values. Here, the macroscopic $\text{Ca}=\mu_\text{o}v/\sigma$ measures the relative importance between viscous and capillary forces. The characteristic velocity is defined as $v=Q/({h}2\pi{r_\text{in}})$, where $r_\text{in}$ is the distance between the cell's center and its closest post. Additionally, for each $f_w$, we design three mixed-wet cells with different base water-wet cluster placements  to verify the reproducibility of the results. Fig.~\ref{fig:phase_diagram} shows the phase diagrams of the fluid-fluid displacement patterns in mixed-wet domains consisting of the three contact angle pairs, and at different Ca and $f_w$. Qualitatively, the displacements display the canonical viscous fingering pattern at high Ca for all mixed-wettability conditions. As Ca decreases, the displacement patterns become more compact with increasing $f_w$, though this effect is much more noticeable in mixed-wet domains with contact angle pair of $30^{\circ}$-$120^{\circ}$ compared to domains with contact angle pairs of $60^{\circ}$-$120^{\circ}$ and $60^{\circ}$-$150^{\circ}$.

\begin{figure*}
	[ht] \centering 
	\includegraphics[width=16.8cm]{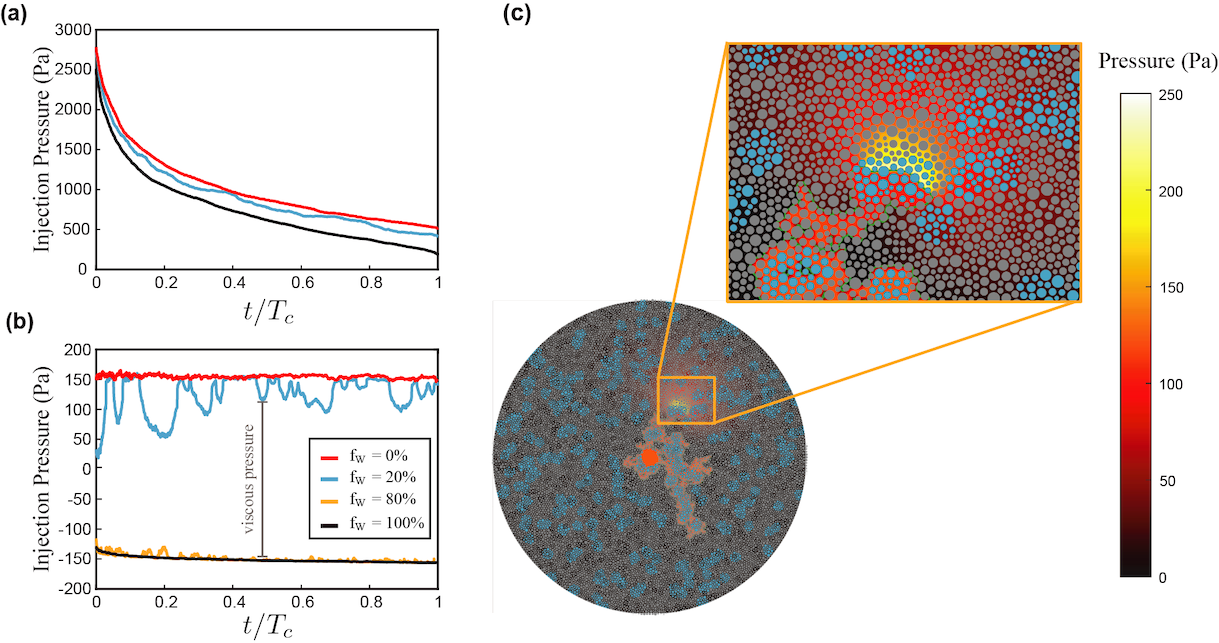} 
	\caption{Evolution of injection pressure as a function of $f_w$ at (a) a high capillary number ($\text{Ca}=5\times10^{-4}$) and (b) a low capillary number ($\text{Ca}=5\times10^{-6}$) for different wettability fractions. The water-wet region has a contact angle of $\theta_\text{w}=30^{\circ}$, while the oil-wet region has a contact angle of $\theta_\text{o}=120^{\circ}$. (c) Snapshot of the pressure map as the invading water fills a water-wet cluster at $\text{Ca}=5\times10^{-6}$ for $f_w=20\%$. The preferential filling of the water-wet cluster causes the local redistribution of the defending oil, leading to non-negligible viscous pressure loss even at very low Ca.} 
\label{fig:Pressure_graphs} 
\end{figure*}

\noindent\textbf{Quantitative measures.} To quantify the morphological properties of the displacement patterns, we calculate the dimensionless finger width ($w_f$) and the fractal dimension ($D_f$). $D_f$ is calculated via the box-counting method~\cite{kenkel1996fractals,schroeder-2009}, while $w_f$ is defined as the average finger width normalized by the average pore size~\cite{primkulov2019signatures,cieplak-prl-1988, cieplakrobbins90,holtzman2016effects}.

Increasing Ca leads to smaller $w_f$ in all mixed-wet domains due to viscous effects. Increasing $f_w$ only yields moderately higher $w_f$ in mixed-wet domains with contact angle pairs of $60^{\circ}$-$120^{\circ}$ and $60^{\circ}$-$150^{\circ}$, though the positive correlation between $w_f$ and $f_w$ is significantly more prominent in mixed-wet domains with a contact angle pair of $30^{\circ}$-$120^{\circ}$ (Fig.~\ref{fig:plots}a). Variations in $w_f$ can be explained by the pore-scale invasion mechanism. The dominant pore invasion mechanism in oil-wet porous media is burst, while pores in water-wet porous media experience more overlap~\cite{primkulov2018quasistatic, primkulov2019signatures}. During overlap events, the invasion of one pore destabilizes the fluid-fluid interface at the neighboring pore, creating a smoother displacement front~\cite{cieplak-prl-1988,cieplakrobbins90,holtzman-prl-2015}. Therefore, $w_f$ increases with increasing $f_w$. Furthermore, the probability of overlap events occurring increases as the invading fluid becomes more wetting to the porous media~\cite{holtzman-prl-2015,primkulov2018quasistatic, primkulov2019signatures}. Indeed, tracking the invasion event type reveals that increase in overlap events with increasing $f_w$ is much more significant in mixed-wet domains with a contact angle pair of $30^{\circ}$-$120^{\circ}$, compared to the mixed-wet domains with contact angle pairs of $60^{\circ}$-$120^{\circ}$ and $60^{\circ}$-$150^{\circ}$ (Fig.~\ref{fig:plots}c-e). 

We measure the fractal dimension of the displacement patterns at the lowest capillary number ($\text{Ca}=5 \times 10^{-6}$), which characterizes the extent to which the displacement patterns fill space in 2 dimensions (Fig.~\ref{fig:plots}b). As expected, $D_f$ increases with increasing $f_w$ in mixed-wet domains with a contact angle pair of $30^{\circ}$-$120^{\circ}$ due to the rise in overlap-driven smoothing of the displacement front. However, this is not the case in mixed-wet domains with contact angle pairs of $60^{\circ}$-$120^{\circ}$ and $60^{\circ}$-$150^{\circ}$, where $D_f$ is a non-monotonic function of $f_w$, and $D_f$ reaches a maximum at $f_w=50\%$. This surprising behavior can be rationalized by the following arguments: at low $f_w$, the broadening of the displacement pattern is controlled by the size and distribution of the water-wet clusters~---~increasing the size and number of water-wet clusters increases $D_f$; at high $f_w$, however, the broadening of the displacement pattern is controlled by the intrinsic wettability of the water-wet clusters~---~more strongly water-wet clusters will yield higher $D_f$ due to a higher probability of overlap events. We note that such contrasting behaviors have been observed in waterflooding experiments in mixed-wet oil-bearing porous media. For instance, Singhal et al.~\cite{singhal1976effect} found that oil recovery efficiency increases monotonically as the fraction of the water-wet surface area of the porous medium increases. In contrast, Skauge~\cite{skauge2002summary, hoiland2007fluid} found that oil recovery efficiency shows a non-monotonic trend according to wettability, with highly mixed-wet cores showing higher oil recovery efficiency than mostly water-wet and oil-wet cores.

\noindent\textbf{Pressure signatures.} We track the injection pressure evolution to gain further insight into the macroscopic impact of mixed-wettability on fluid-fluid displacement in porous media. Injection pressure is a valuable piece of information in various subsurface applications including geological carbon sequestration~\cite{bachu-pecs-2008} and hydraulic fracturing~\cite{warner-pnas-2012}, and it can help inform the wettability state of the porous medium~\cite{sygouni2006capillary}. The injection pressure consists of the capillary pressure across the interface between the invading and defending fluid, and the combined viscous pressure loss in the two fluids.

At high Ca, viscous pressure loss dominates and the injection pressure decreases monotonically as the more viscous defending fluid is pushed out by the less viscous invading fluid. This decreasing trend in injection pressure is observed in all cases regardless of the wettability state of the domain~(Fig.~\ref{fig:Pressure_graphs}a). Interestingly, the signature of wettability is evident even at the highest capillary number ($\text{Ca}=5\times10^{-4}$)~---~the injection pressure is consistently higher in a purely oil-wet domain compared to a purely water-wet domain. These results agree with the injection pressure measurements in uniform-wet microfluidic experiments~\cite{zhao-pnas-2016}. The injection pressures observed in mixed-wet domains fall between those in purely oil-wet and water-wet domains~(Fig.~\ref{fig:Pressure_graphs}a).

At low Ca, capillary pressure dominates and the injection pressure is controlled by the wettability of the domain. Indeed, the injection pressure fluctuates around a mean \emph{positive} value in a purely oil-wet domain (i.e., drainage), but around a mean \emph{negative} value in a purely water-wet domain (i.e., imbibition). Closer inspection shows that the injection pressure fluctuations are larger in a purely oil-wet domain than in a purely water-water domain~(Fig.~\ref{fig:Pressure_graphs}b, c). This difference in fluctuations is attributed to the prevalence of burst invasion events in drainage, whose critical capillary pressures are larger than those of touch and overlap invasion events, which are more prevalent in imbibition~\cite{maaloy1992dynamics,moebius-jcis-2012,primkulov2019signatures}. 

Intuitively, at vanishingly small Ca, one would expect the injection pressure to fluctuate between the drainage capillary pressure and the imbibition capillary pressure in a mixed-wet domain, as the invading water transits oil-wet and water-wet clusters. Surprisingly, this is not what we see in mixed-wet domains with small $f_w$~---~instead, we observe that the injection pressure fluctuates between the drainage capillary pressure and some pressure that is much higher than the imbibition capillary pressure~(Fig.~\ref{fig:Pressure_graphs}b). This deviation can be explained by the fact that once the invading water encounters a water-wet cluster, all of the injected volume will preferentially enter that region of the micromodel. Since the water-wet clusters are small at low $f_w$, the preferential filling within this localized region will cause the defending fluid to redistribute along the invasion front, leading to non-negligible viscous pressure loss even at very low Ca. Indeed, as the invading water enters a water-wet cluster ($\theta_\text{w}=30^{\circ}$, $f_w=20\%$), the difference between the measured injection pressure and the imbibition capillary pressure equal the viscous pressure in the defending oil~(Fig.~\ref{fig:Pressure_graphs}c). Therefore, mixed-wettability could decrease the effective permeability of porous media at low Ca. As the invading water exits the water-wet cluster, the injection pressure increases to the drainage capillary pressure. Consequently, the period of the fluctuations indicates the size of the water-wet cluster. At high $f_w$, the invading water has access to many connected water-wet pores at any given time, and the injection pressure closely tracks the imbibition capillary pressure~(Fig.~\ref{fig:Pressure_graphs}b).

%The oil and gas industry spends significant efforts on core analysis to determine rock properties such as porosity, permeability, rela- tive permeability, and wettability, as a part of subsurface workflows (McPhee et al. 2015). Fundamental to these analyses is relative permeability, which measures the ease at which one phase flows in the presence of another, immiscible phase (Bear 2018, Bear 1970).

%There are no simple rules that quantify the wide range of observed relative permeability behaviors despite great efforts to find them (Blunt 2017). Experimental and simulation data often provide conflicting results, such as trends of residual oil saturation and endpoint relative permeability versus wettability (Fan et al. 2020, Christensen and Tanino 2019, Jadhunandan and Morrow 1995).

\begin{figure*}
	[htp] \centering 
	\includegraphics[width=16.8cm]{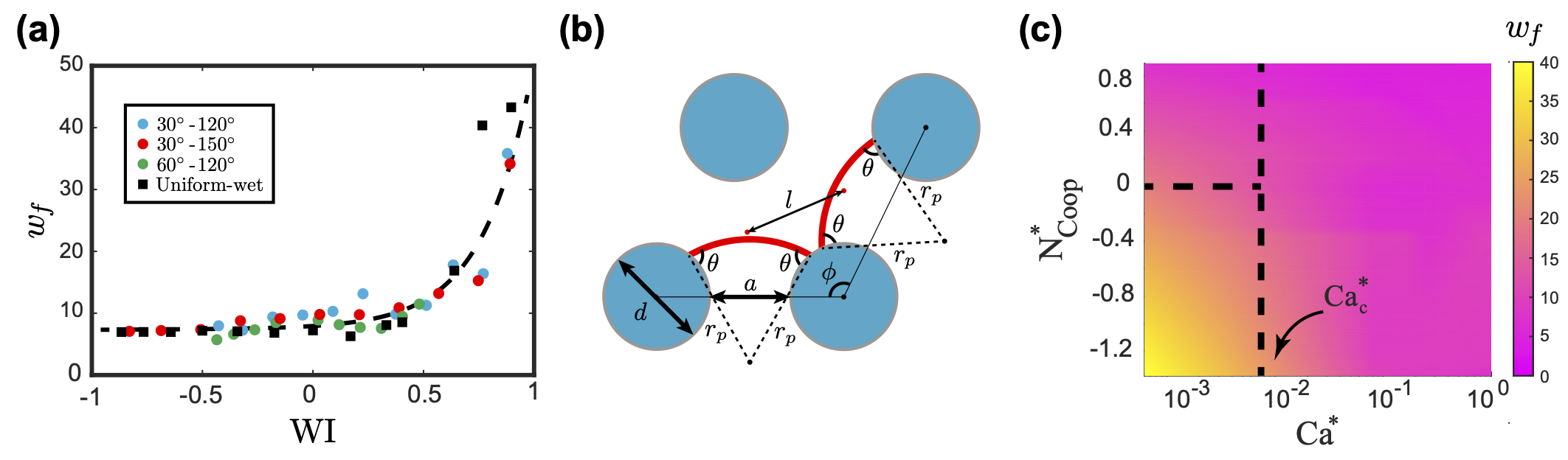} 
	\caption{Wettability index description of mixed-wet systems. (a) We extract the average finger width $w_f$ from fluid-fluid displacement simulations in more than 50 mixed-wet porous media of different contact angle pairs ($\theta_\text{w}$-$\theta_\text{o}$$=30^{\circ}$-$120^{\circ}$, $60^{\circ}$-$120^{\circ}$, $60^{\circ}$-$150^{\circ}$) and wettability fractions. Plotting $w_f$ vs the wettability index (WI) collapses the dataset onto a single curve, which matches the results obtained from simulations performed in uniform-wet porous media with different wettabilities. $\text{Ca}=5\times10^{-6}$ in all simulations. (b) Invasion front configuration at two adjacent pores. (c) $\text{Ca}^{\star}$ and $N_\text{coop}^{\star}$ explain the variations in $w_f$ in 15 different mixed-wet domains ($\theta_\text{w}$-$\theta_\text{o}$$=30^{\circ}$-$120^{\circ}$, $60^{\circ}$-$120^{\circ}$, $60^{\circ}$-$150^{\circ}$; $f_\text{w}=2\%, 26\%, 50\%, 76\%, 98\%$) over a wide range of capillary numbers.} 
\label{fig:scaling_analysis} 
\end{figure*}

\noindent\textbf{Wettability index description of mixed-wet systems.} Despite the complexities involved in fully characterizing the wettability of mixed-wet systems (e.g., contact angles, wettability fraction, etc.), it is sometimes helpful to approximate the wettability state of a porous medium by a single parameter. One of the simplest representations of this idea is the wettability index (WI), which in its general form is given by~\cite{armstrong2021multiscale}
\begin{equation}
	\text{WI}=\sum_{n=1}^{i} \frac{\sigma_{i1}-\sigma_{i2}}{\sigma_{ \text{12}}} f_i, 
\label{eq:eight} \end{equation}
where $f_i$ is the areal fraction of surface $i$, $\sigma_{12}$ is the interfacial tension between fluid 1 and fluid 2, and $\sigma_{i1}$, $\sigma_{i2}$ are the interfacial tensions between surface $i$ and the two fluids. Young's equation relates the interfacial tensions to the contact angle. For our mixed-wet system with two distinct contact angles and oil-water as the fluid-fluid pair, incorporating Young's equation gives
\begin{equation}
	\text{WI}= f_\text{w} \cos \theta_\text{w} +(1-f_\text{w}) \cos \theta_\text{o}. 
\end{equation}
Therefore, WI is a weighted average description of the overall wettability of the porous media. Here, we examine the effectiveness of WI in predicting the fluid-fluid displacement pattern as characterized by the finger width $w_f$ at low capillary number ($\text{Ca}=5\times10^{-6}$). We find that WI successfully collapses $w_f$ extracted from simulations conducted in more than 50 different mixed-wet porous media. Additionally, this master curve matches the one obtained from simulations of fluid-fluid displacement in uniform-wet porous media, which both show that finger width increases with increasing WI (Fig.~\ref{fig:scaling_analysis}a). 

We take the weighted average description of fluid-fluid displacement in mixed-wet porous media one step further via scaling analysis and derive two dimensionless parameters that capture the behavior of $w_f$ across all Ca. We first compare the relative importance of the characteristic viscous pressure $\delta p_v$ and capillary pressure $\delta p_c$ at the pore-scale~\cite{holtzman2016effects,primkulov2019signatures,holtzman-pre-2010, toussaint2005influence}. The characteristic viscous pressure is given by the pressure drop in the viscous defending fluid over a characteristic length $l$ 
\begin{equation}
	\delta p_v=32~\mu_\text{def}{v}\frac{(a+h)^2}{a^2 h^2}l, 
\label{eq:nine} \end{equation}
where $v$ is the characteristic injection velocity, $a$ and $l$ are the median pore length and pore throat size, respectively. The characteristic capillary pressure is assumed to be the capillary pressure corresponding to a burst event at a characteristic pore throat 
\begin{subequations}
	\begin{equation}
		\delta{p_c} = \sigma\kappa, 
	\end{equation}
	\begin{equation}
		\kappa=\frac{2\cos \theta}{h} + \frac{2}{{a}(\hat{l} \cos \theta + \sqrt{\hat{l}^2 {\cos \theta}^2 + 1 + 2\hat{l} })}, 
	\label{eq:kappa} \end{equation}
\end{subequations}
where $\kappa$ is the critical interface curvature required for burst to occur (detailed derivation of Eq.~\ref{eq:kappa} is included in the \emph{Supplemental Material}), $\hat{l}={d}/{a}$ is the ratio between the median post diameter and the median pore throat size (Fig.~\ref{fig:scaling_analysis}b). The ratio between $\delta p_v$ and $\delta p_c$ yields a modified capillary number $\text{Ca}^{\star}$, which takes into account the wettability of the system~\cite{holtzman-prl-2015,primkulov2019signatures,holtzman2016effects}. In our mixed-wet system, water-wet and oil-wet regions have the same characteristic viscous pressure drop, but noticeably different capillary pressure. We take the weighted average of the characteristic capillary pressures in the water-wet and oil-wet regions and arrive at 
\begin{equation}
	\text{Ca}^{\star} = \frac{32(a+h)^2 l}{a^2 h^2 (f_\text{w} \kappa_\text{w} +(1-f_\text{w}) \kappa_\text{o})}\text{Ca}, 
\end{equation}
where $\kappa_\text{w}$ and $\kappa_\text{o}$ are the critical interface curvatures (Eq.~\ref{eq:kappa}) for water-wet and oil-wet regions, respectively.

$\text{Ca}^{\star}$ alone is not able to capture the displacement pattern across different wettabilities, since it does not account for the different types of invasion events (i.e., burst, touch, overlap). It is well-known that overlap events promote cooperative pore filling, which leads to more compact displacement (i.e. wider fingers)~\cite{cieplak-prl-1988,cieplakrobbins90,zhao-pnas-2016,primkulov2018quasistatic}. To capture the effect of cooperative pore filling, we calculate the dimensionless cooperative number ($N_\text{coop}$) for our system. Specifically, $N_\text{coop}$ evaluates the relative likelihood of overlap and burst events by comparing the critical capillary pressures associated with the two invasion event types~\cite{holtzman-prl-2015}. $N_\text{coop}$ for a characteristic pore is given by 
\begin{subequations}
	\begin{equation}
		N_\text{coop}= \frac{\phi}{2}+ \theta + \arccos({\frac{r_{p}^2 + d_1^2 - \frac{d^2}{4}}{2 d_1 r_{p}}}) + \arccos({\frac{ a + d }{2 d_1}})- \pi ,
	\end{equation}
	\begin{equation}
		r_{p}= \frac{\hat{l} \cos \theta + \sqrt{\hat{l}^2 {\cos \theta}^2 +1 + 2\hat{l} }}{\frac{2}{ a}},
	\end{equation}
\end{subequations}
where $\phi$ is the median angle between two neighboring interfaces. Detailed derivation of $N_\text{coop}$ is included in the \emph{Supplemental Material}. Positive $N_\text{coop}$ indicates burst events are more likely to occur than overlap, while negative $N_\text{coop}$ suggests the opposite. Finally, we take the weighted average of the cooperative number for the water-wet region ($N_\text{coop,w}$) and the oil-wet region ($N_\text{coop,o}$) to arrive at a modified cooperative number for the mixed-wet system
\begin{equation}
	N_\text{coop}^{\star}= f_\text{w}N_\text{coop,w}+(1-f_\text{w})N_\text{coop,o}.
\end{equation}
$\text{Ca}^{\star}$ and $N_\text{coop}^{\star}$ explain the variations in $w_f$ in different mixed-wet domains over a wide range of capillary numbers (Fig.~\ref{fig:scaling_analysis}c). In our system, the critical modified capillary number $\text{Ca}^{\star}_c\sim{d/D}\approx{6\times10^{-3}}$, where $D$ is the diameter of the domain. For $\text{Ca}^{\star}>\text{Ca}^{\star}_c$, viscous pressure dominates, and the displacement pattern consists of long, thin fingers (small $w_f$) that are typical in viscous fingering. For $\text{Ca}^{\star}<\text{Ca}^{\star}_c$, the displacement pattern is dependent upon the amount of cooperative smoothing in the system such that wide fingers (large $w_f$) start to emerge for $N_\text{coop}^{\star}<0$. 

\section{Conclusions}

We have investigated fluid-fluid displacement in simple mixed-wet porous media consisting of distinct water-wet and oil-wet regions using a dynamic pore network model. We generate phase diagrams of the displacement patterns for three different water-wet/oil-wet contact angle pairs over a wide range of $\text{Ca}$~(Fig.~\ref{fig:phase_diagram}). The impact of mixed-wettability is most prominent at low $\text{Ca}$, whose effect on the displacement pattern is controlled by the complex interplay between wettability fraction and the intrinsic contact angle of the water-wet regions~(Fig.~\ref{fig:plots}). Our simulations also provide insights into the pressure signature of fluid-fluid displacement in mixed-wet porous media, which displays fluctuations that cannot be explained by capillary pressure alone, even at vanishingly small $\text{Ca}$~(Fig.~\ref{fig:Pressure_graphs}). We find that the injection pressure fluctuations are modulated by the viscous pressure of the defending fluid due to preferential filling of isolated water-wet regions, and the duration of the fluctuations are determined by the size of the water-wet regions. One surprising consequence of this complex interplay is that at low Ca, mixed-wettability could simultaneously result in reduced effective permeability, but more effective sweep of the defending fluid. Finally, we derive scaling arguments based on a weighted average description of the overall wettability of the mixed-wet porous medium, which effectively capture the variations in displacement pattern morphology as characterized by the finger width~(Fig.~\ref{fig:scaling_analysis}). Our study presents a systematic understanding of the relationship between contact angle, wettability fraction, and capillary number in governing fluid-fluid displacement in simple mixed-wet porous media, and it serves as a platform upon which more complex mixed-wet porous media can be investigated.

\section{Appendix A. Model geometry}

In our system, the diameters of the posts follow a Gaussian-like distribution that ranges from 220 to 1700~$\mu$m. Additionally, the pore-throat size follows a lognormal-like distribution that ranges from 50 to 700~$\mu$m. 

\begin{figure*}[hbp]
	\centering
	\includegraphics[width=16cm]{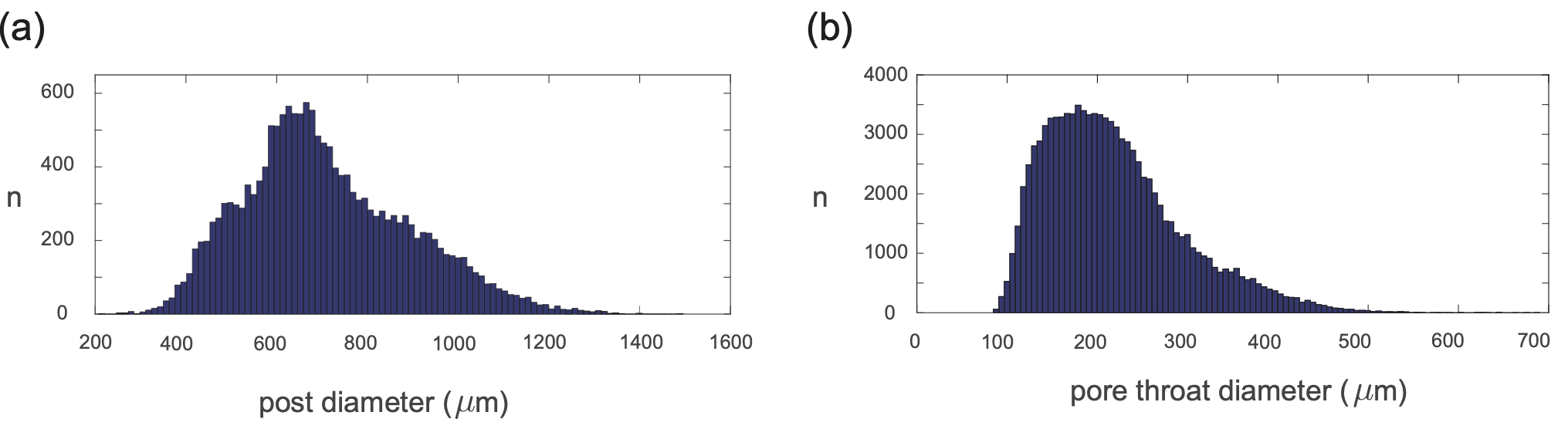}
	\caption{Characteristics of the micromodel domain: (a) post diameter (b) pore-throat size.}\label{fig:sample}
\end{figure*}

\section{Appendix B. Critical interface curvature for burst}

\begin{figure*}[hbp]
	\centering
	\includegraphics[width=10cm]{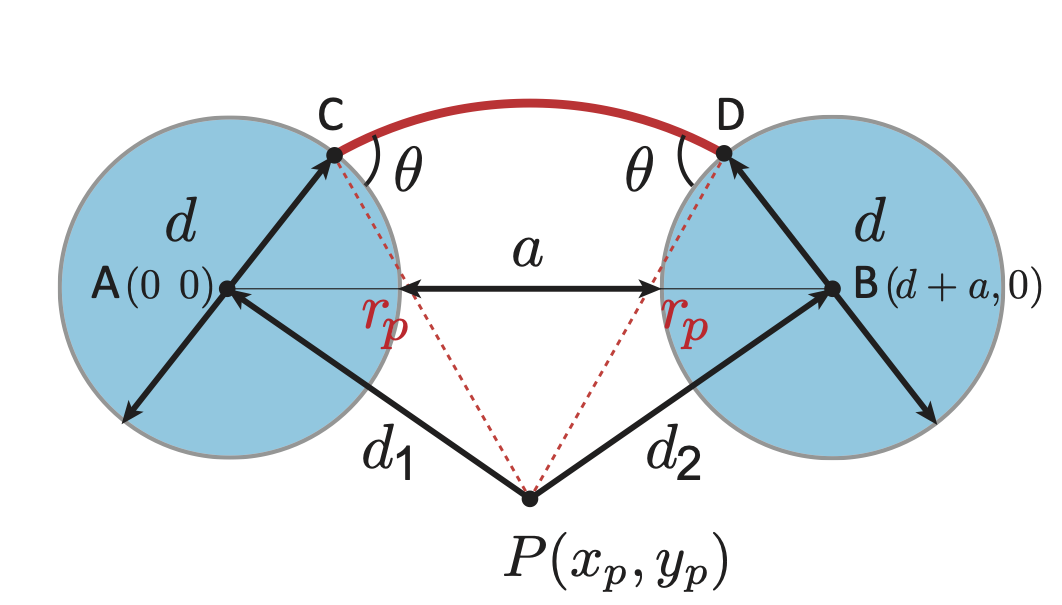}
	\caption{Fluid-fluid interface configuration at a median size pore throat.}\label{fig:burst_curvature}
\end{figure*}

Our aim here is to determine the critical interface curvature for a burst to occur in a typical pore throat configuration in the model. In this configuration, the fluid-fluid interface connects two posts with diameters $d$ (i.e., the median post diameter in the system) where the closest distance between these posts is equal to $a$ (i.e., the median pore throat size in the system). The wettability of the posts is the same and their contact angles are equal to $\theta$. After translation and rotation of the coordinate system, the center of post A is positioned at $(0,0)$ and the center of post B is located at (${a+d}, 0$).  Fig.~\ref{fig:burst_curvature} shows the fluid-fluid interface arc $\overset{\huge\frown}{CD}$ with the center of $P(x_p,y_p)$ in the pore throat. We denote the distance between points A and P as $d_1$, which is given by the law of cosine at the triangle ${\triangle}{APC}$ as
\begin{equation}
	d_1 =\sqrt {\left(\frac{d}{2}\right)^2+r_p^2 - 2 \left(\frac{d}{2}\right) r_p \cos \theta}. \label{eq:onea}
\end{equation}
We denote the distance between points B and P as $d_2$, which is equal to $d_1$.

As defined in \citet{primkulov2018quasistatic}, a burst event occurs when the center of the fluid-fluid interface has the following configuration
\begin{equation}
X_p^2=d_1^2,
\label{eq:onec}
\end{equation}
\begin{equation}
(X_p-X_2)^2=d_2^2, 
\label{eq:oned}
\end{equation}
where $X_2$ is the $x$ coordinate of the center of post B, and it is equal to $a+d$. Solving equations \ref{eq:onea}-\ref{eq:oned}, we find the radius of curvature of the fluid-fluid interface corresponding to a burst event as
\begin{equation}
r_p=\frac{\hat{l} \cos {\theta} + \sqrt{\hat{l}^2 (\cos \theta)^2 + 1 + 2\hat{l} }}{\frac{2}{a}},
 \label{eq:onee}
\end{equation}
where $\hat{l}=d/a$. The mean curvature of the interface at a burst event is defined as $\kappa=(1/r_p+1/r_\text{out})$, where $r_\text{out}=h/(2\cos\theta)$ is the out of plane curvature of the interface. Consequently, the critical interface curvature for a burst to occur is equal to
\begin{equation}
	\kappa=\frac{2\cos \theta}{h} + \frac{2}{{a}(\hat{l} \cos \theta + \sqrt{\hat{l}^2 {\cos \theta}^2 + 1 + 2\hat{l} })}.
	 \label{eq:onef}
\end{equation}

\section{Derivation of the dimensionless cooperative number}

\begin{figure*}[hbp]
	\centering
	\includegraphics[width=10cm]{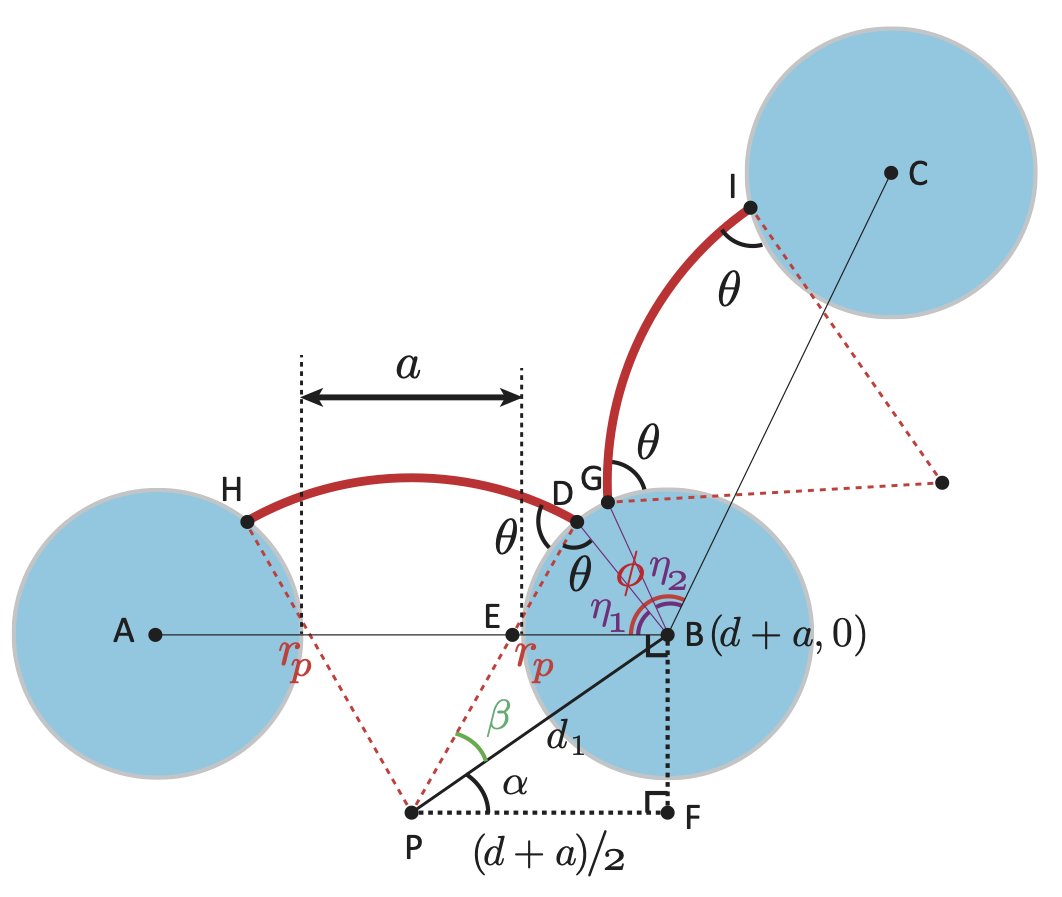}
	\caption{Fluid-fluid interface configuration at a median size pore throat.
	}\label{fig:N_coop}
\end{figure*}

To determine whether a burst event precedes an overlap event in a pore throat, we define the dimensionless cooperative number $N_\text{coop}$. Fig.~\ref{fig:N_coop} shows two adjacent pore throats consisting of three posts with similar diameter $d$, pore throat size $a$ and contact angle $\theta$ which is the most representative for our system. The angle between two neighboring interfaces $\angle ABC$ is denoted $\phi$. We denote the angle between the line connecting points D and B and the line connecting A and B as $\eta_1$. Similarly, we denote the angle between the line connecting point G and B and the line connecting B and C as $\eta_2$. 

Increasing the pressure of the invading fluid advances the two adjacent interfaces within the pore throat until they meet (i.e., an overlap event occurs). These two interfaces must satisfy the geometric condition $\eta_1+\eta_2=\phi$~\cite{Holtzman2015}. To determine whether a burst event is preceded by an overlap event, we must calculate $\eta_1$ and $\eta_2$ corresponding to a burst event, and then calculate $\lambda$
\begin{equation}
\lambda= \phi-(\eta_1+ \eta_2).
\label{eq:twoa}
\end{equation}
Here, $\lambda>0$ indicates that $\eta_1+\eta_2<\phi$, and thus a burst occurs before the interfaces can overlap. In contrast, $\lambda<0$ suggests that $\eta_1+\eta_2>\phi$, which means that an overlap event will take place and the interfaces merge and form a new fluid-fluid interface. Given that the pore throats in the system are similar, and consequently the burst radius of the interfaces, we can rewrite the equation with $\eta_1=\eta_2=\eta$ and define the cooperative number $N_\text{coop}$ as
\begin{equation}
N_\text{coop}=\frac{\lambda}{2}= \frac{\phi}{2}-\eta.
 \label{eq:twob}
\end{equation}

\begin{figure*}[bp]
 	\centering
 	\includegraphics[width=18cm]{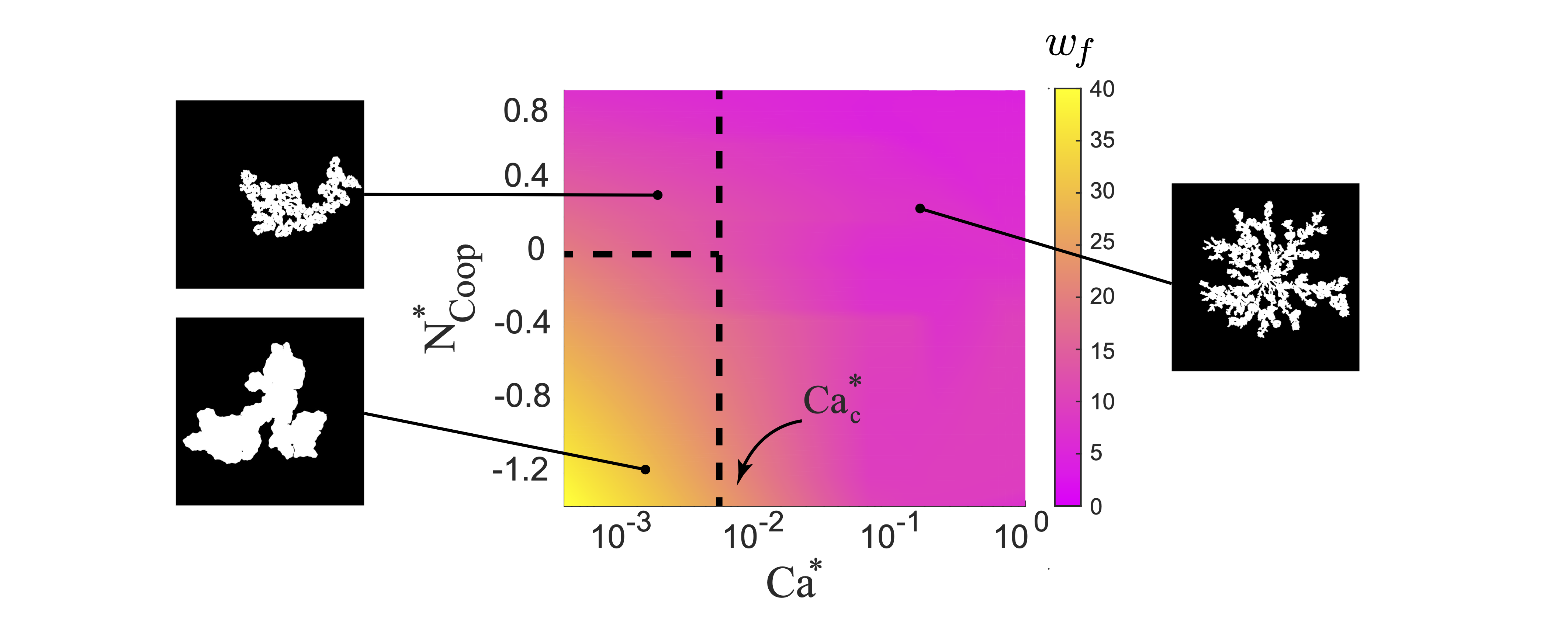}
 	\caption{$\text{Ca}^{\star}$ and $N_\text{coop}^{\star}$ explain the variations in $w_f$ in 15 different mixed-wet domains ($\theta_\text{w}$-$\theta_\text{o}$$=30^{\circ}$-$120^{\circ}$, $60^{\circ}$-$120^{\circ}$, $60^{\circ}$-$150^{\circ}$; $f_\text{w}=2\%, 26\%, 50\%, 76\%, 98\%$) over a wide range of capillary numbers.}\label{fig:burst_curvature}
\end{figure*}

In Fig.~\ref{fig:N_coop}, we connect points P and B to form a line with length $d_1$. We then define the right ${\triangle}{BPF}$ with acute angle $\alpha$ given by
\begin{equation}
\alpha =\arccos \left(\frac{{a+d}}{2 d_1}\right).
 \label{eq:twoc}
\end{equation}
Applying the law of cosines in the triangle ${\triangle}{BPD}$ gives
\begin{equation}
\frac{d}{2}=\sqrt {r_p^2+d_1^2-2 d_1 r_p \cos {\beta}},
 \label{eq:twod}
\end{equation}
where $\beta=\pi-\eta-\alpha-\theta$. We rearrange Eq.~\ref{eq:twod}
\begin{equation}
\eta = \pi - \arccos (\frac{r_p^2+d_1^2-\frac{d^2}{4}}{2 d_1 r_p})- \arccos \frac{a+d}{2 d_1}- \theta \label{eq:twoe}.
\end{equation}
Substituting Eq.~\ref{eq:twoe} into Eq.~\ref{eq:twob} gives
\begin{equation}
	N_\text{coop}= \frac{\phi}{2}+ \theta + \arccos({\frac{r_{p}^2 + d_1^2 - \frac{d^2}{4}}{2 d_1 r_{p}}}) + \arccos({\frac{ a + d }{2 d_1}})- \pi. \label{eq:twof}
\end{equation}

%%%%%%%%%%%%%%%%%%%%%%%%%%%%%%%%%%%%%%%%%%%%%%%%%%%%%%%%%%%%
\bibliography{characteristics_mixwet}
%%%%%%%%%%%%%%%%%%%%%%%%%%%%%%%%%%%%%%%%%%%%%%%%%%%%%%%%%%%%
\end{document}